\documentclass[prl,floatfix,showpacs,amsmath,twocolumn]{revtex4}
\usepackage{amssymb}
\usepackage{graphics}

\newcommand{\be}{\begin{eqnarray}}
\newcommand{\ee}{\end{eqnarray}}
\newcommand{\bea}{\begin{eqnarray}}
\newcommand{\eea}{\end{eqnarray}}
\newcommand{\bma}{\begin{subequations}}
\newcommand{\ema}{\end{subequations}}

\begin{document}

\title{Renormalization algorithms for Quantum-Many Body Systems in
two and higher dimensions}

\author{F. \surname{Verstraete}}
\affiliation{Max-Planck-Institut f\"ur Quantenoptik,
Hans-Kopfermann-Str. 1,
  Garching, D-85748, Germany.}
\author{J. I. \surname{Cirac}}
\affiliation{Max-Planck-Institut f\"ur Quantenoptik,
Hans-Kopfermann-Str. 1,
  Garching, D-85748, Germany.}

\pacs{PACS}
\date{\today}

\begin{abstract}
We describe quantum many--body systems in terms of projected
entangled--pair states, which naturally extend matrix product
states to two and more dimensions. We present an algorithm to
determine correlation functions in an efficient way. We use this
result to build powerful numerical simulation techniques to
describe the ground state, finite temperature, and evolution of
spin systems in two and higher dimensions.
\end{abstract}

\maketitle

The theoretical investigation of strongly correlated systems is
one the most challenging tasks in several fields of Physics. Even
though several analytical techniques and numerical methods have
been developed during the last decades, there still exist a rich
variety of systems which remain untractable. Even some of the
simplest systems, which deal with spins in lattices with short
range interactions, are very hard to simulate numerically. The
development of powerful numerical techniques would allow us to
discover a rich variety of intriguing phenomena and to confirm
some of the predictions which have been made.

In the case of 1D systems, much analytical insight has been gained
by finding exact expressions for the ground and/or excited
eigenstates of some particular Hamiltonians, as it is the case for
the 1D--AKLT states \cite{AKLT}. On the other hand, a very
powerful numerical simulation method known as DMRG \cite{White}
has allowed us to determine physical properties of generic spin
chains to an unprecedented accuracy. Recent work has also shown
how DMRG can be adapted to simulate the spin dynamics at
zero-temperature \cite{VidalDaleyWhite} or at finite temperature
and in the presence of dissipation \cite{Zwolak,VGC04}. The
success of the DMRG method and its extensions relies on the
existence of the so-called matrix product states (MPS)
\cite{Fannes}, which generalize the 1D--AKLT states. The DMRG
method can be understood as a variational method within these MPS
\cite{Ostlund,Dukelsky,VPC04}, and part of its success relays on
the fact that correlation functions can be efficiently calculated.

In two or higher dimensions, however, almost no models have been
solved exactly. A generalization of DMRG to higher dimensions is
hard to scale, as the MPS--ansatz explicitly assumes a 1D
configuration. The Monte Carlo method \cite{MC}, on the other
hand, is very useful to describe certain systems, but for models
with frustration is, unlike DMRG, plagued by the so-called sign
problem. The physics of 2D spin systems is therefore not very well
understood as compared to 1D systems; this is very unfortunate as
a good understanding would shed new light on many open questions
in condensed matter, such as high-$T_c$ superconductivity.

In this paper, we present a natural generalization of the 1D MPS
to two and higher dimensions and build simulation techniques based
on those states which effectively extend DMRG to higher
dimensions. We call those states {\em projected entangled--pair
states} (PEPS), since they can be understood in terms of pairs of
maximally entangled states of some auxiliary systems, and that are
projected in some low--dimensional subspaces locally. This class
of states includes the generalizations of the 2D AKLT-states known
as tensor product states \cite{Nishino04} which have been used for
2D problems (see also \cite{Niggeman,Martindelgado,VC03}) but is
much broader since every state can be represented as a PEPS (as
long as the dimension of the entangled pairs is large enough). We
also develop an efficient algorithm to calculate correlation
functions of these PEPS, and which allows us to extend the 1D
algorithms \cite{White,VidalDaleyWhite,VGC04,Zwolak} to higher
dimensions. This leads to many interesting applications, such as
scalable variational methods for finding ground or thermal states
of spin systems in higher dimensions as well as to simulate their
time-evolution. For the sake of simplicity, we will present our
results for a square lattice in 2D, but they are easily
generalized to higher dimensions and other geometries.

\begin{figure}[t]
  \centering
  \resizebox{\linewidth}{!}{\includegraphics{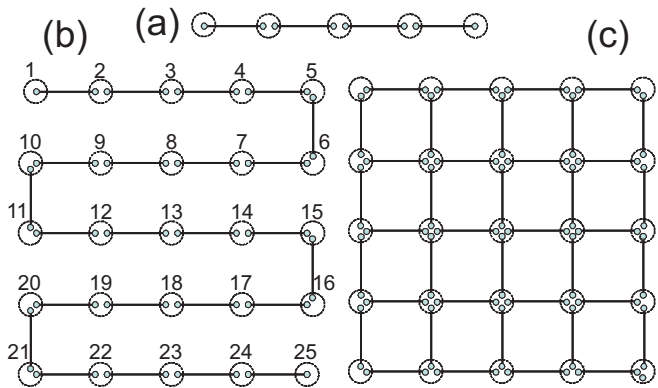}}
  \caption{Graphical representation of MPS in 1 dimension
  (a), in 2 dimensions (b), and of PEPS (c). The bonds represent
  pairs of maximally entangled D--dimensional auxiliary spins and the
  circles denote projectors that map the inner auxiliary spins
  to the physical ones.}
  \label{figscheme}
\end{figure}

Let us start by recalling the representation introduced in
\cite{VPC04} of the state $\Psi$ of $N$ $d$--dimensional systems
in terms of MPS. For that, we substitute each physical system $k$
by two auxiliary systems $a_k$ and $b_k$ of dimension $D$ (except
at the extremes of the chain). Systems $b_k$ and $a_{k+1}$ are in
a maximally entangled state $|\phi\rangle=\sum_{n=1}^D
|n,n\rangle$, which is represented in Fig.\ \ref{figscheme}(a) by
a solid line (bond) joining them. The state $\Psi$ is obtained by
applying a linear operator $Q_k$ to each pair $a_k,b_k$ that maps
the auxiliary systems onto the physical systems, i.e.
 \bea
\nonumber
 |\Psi\rangle &=& \hspace{1cm}Q_1\otimes Q_2\otimes\ldots Q_N
 \hspace{.2cm}|\phi\rangle \ldots |\phi\rangle
 \\
 \label{MPS}
 &=& \sum_{s_1,\ldots,s_N=1}^d {\rm F}_1
 (A^{s_1}_1, \ldots, A^{s_N}_N) |s_1,\ldots,s_N\rangle,
 \eea
where the matrices $A^s_k$ have elements $(A^s_k)_{l,r}=\langle
s|Q_k|l,r\rangle$. Note that the indices $l$ and $r$ of each
matrix $A^s_k$ are related to the left and right bonds of the
auxiliary systems with their neighbors, whereas the index $s$
denotes the state of the physical system. The function F$_1$ is
just the trace of the product of the matrices, i.e. it contracts
the indices $l,r$ of the matrices $A$ according to the bonds.

As shown in \cite{VPC04}, every state can be represented in the
form (\ref{MPS}) as long as the dimension $D$ can be chosen
sufficiently large. Note, however, that the above picture of the
state is basically one dimensional, since each auxiliary system is
entangled only to one nearest neighbor. Thus, these states appear
to be better suited to describe 1D systems, with short range
interactions, since a small local dimension $D$ may give a good
approximation to the real state of the whole system. Note also
that, as it is clear from the above representation, any block of
systems is only entangled to the rest by at most two maximally
entangled state of the auxiliary particles and thus its entropy is
bounded by $2\log_2 D$, independent of the block size. This has
been identified as the main reason why DMRG does not describe well
critical systems, where the entropy grows with the logarithm of
the block size \cite{Vidallog}.

States in the form (\ref{MPS}) have also been used to represent 2D
systems \cite{Xiang}. For simplicity let us thus consider a 2D
square lattice of $N:=N_h\times N_v$ systems. The idea there is to
numerate them in such a way that they can be regarded as a long 1D
system [Fig.\ \ref{figscheme}(b)]. In general, this method cannot
be extended to larger systems since nearest neighbor interactions
in the original 2D system (for example between $11$ and $20$) give
rise to long interactions in the effective 1D description.
Moreover, the entropy of some blocks does not scale as the area of
the block, as it is expected for 2D configurations. For example,
the block formed by systems from $6$--$15$ has at most a constant
entropy of $2\log_2 D$.

For 2D systems we propose to use the description based on Fig.
\ref{figscheme}(c). Each physical system of coordinates $(h,v)$ is
represented by four auxiliary systems $a_{h,v}$, $b_{h,v}$,
$c_{h,v}$, and $d_{h,v}$ of dimension $D$ (except at the borders
of the lattices). Each of those systems is in a maximally
entangled state $\phi$ with one of its neighbor, as shown in the
figure. The state $\Psi$ is obtained by applying to each site one
operator $Q_{h,v}$ that maps the auxiliary systems onto the
physical systems:
 \be
 \label{repres2D}
 |\Psi\rangle
 = \sum_{s_{1,1},\ldots,s_{N_h,N_v}=1}^d {\rm F}_2
 (\{A^{s_{h,v}}_{h,v}\} ) |s_{1,1},\ldots,s_{N_h,N_v}\rangle.
 \ee
Here, the $A$'s are four index tensors with elements
$(A^s_{h,v})_{u,d,l,r}=\langle s|Q_{h,v}|u,d,l,r\rangle$. As in
the 1D case, we associate each index of such tensors to each
direction (up, down, left, and right). Thus, the position with
coordinates $(h,v)$ is represented by a tensor
$(A^s_{h,v})_{u,d,l,r}$ whose index $s$ is represents the physical
system, and the other four indices are associated with the bonds
between the auxiliary systems and the neighboring ones. The
function F$_2$ contracts all these indices $u,d,l,r$ of all
tensors $A$ according to those bonds. Note that we can generalize
this construction to any lattice shape and dimension, and that
using the construction of \cite{VPC04} one can show that any state
can be written as a PEPS. In this way, we also resolve the problem
of the entropy of blocks mentioned above, since now this entropy
is proportional to the bonds that connect such block with the
rest, and therefore to the area of the block. Note also that, in
analogy to the MPS \cite{Fannes}, the PEPS are guaranteed to be
ground states of local Hamiltonians.

We show now how to determine expectation values of operators in
the state $\Psi$ (\ref{repres2D}). We consider a general operator
$O=\prod_{h,c} O_{h,c}$ and define the four--indices tensor
 \be
 (E_{O_{h,c}})_{\tilde u,\tilde d,\tilde l,\tilde r}:=
 \sum_{s,s'=1}^d \langle
 s|O_{h,c}|s'\rangle
 (A^s)_{u,d,l,r} (A^{s'})^\ast_{u',d',l',r'}
 \ee
where the indices are combined in pairs, i.e., $\tilde u=(u,u'),
\tilde d=(d,d'), \tilde l=(l,l')$, and $\tilde r=(r,r')$. One can
easily show that $\langle \Psi|O|\Psi\rangle={\rm
F}_2(E_{O_{h,c}})$. Thus, the evaluation of expectation values
consists of contracting indices of the tensors $E$. In order to
show how to do this in practice, we notice that the tensors
associated to the first and last rows, once contracted, can be
reexpressed in terms of a MPS. In particular, we define [compare
(\ref{MPS})]
 \bma
 \bea
 \label{MPS2}
 |U_1\rangle &:=& \sum_{d_1\ldots d_N=1}^{D^2} {\rm F}_1
 (E^{d_1}_{1,1} \ldots E^{d_N}_{1,N}) |d_1\ldots d_N\rangle,\\
 \langle U_{N_v}| &:=& \sum_{u_1\ldots u_N=1}^{D^2} {\rm F}_1
 (E^{u_1}_{N,1} \ldots E^{u_N}_{N,N}) \langle u_1\ldots u_N|.\label{MPS3}
 \eea
 \ema
Here we have used the short--hand notation $E_{h,c}:=E_{O_{h,c}}$,
and the fact that the tensors in the first and last rows have at
most three indices [see Fig.\ \ref{figscheme}c]. Thus, the
horizontal indices $(l,r)$ of the tensors play the role of the
indices of each matrix, whereas the vertical ones $(d)$ plays the
role of the indices corresponding to the physical systems in 1D.
Analogously, the rows $2,3,\ldots, N_v-1$ can be considered as
matrix product operators (MPO) \cite{VGC04},
 \[
 U_k:=\sum_{d_1,u_1\ldots=1}^{D^2} {\rm F}
 (E^{d_1,u_1}_{1,1}\hspace{-.3cm},\ldots,E^{d_N,u_N}_{1,N}) |d_1\ldots d_N\rangle
 \langle u_1\ldots u_N|.
 \]
We have $\langle \Psi|O|\Psi\rangle= \langle U_N|U_{N-1}\ldots
U_2|U_1\rangle$.

The evaluation of expectation values poses a serious problem since
the number of indices proliferate after each contraction. For
example, the vector $|U_2\rangle:=U_2|U_1\rangle$ can be written
as the MPS (\ref{MPS2}) but with the substitution
 \be
 E^{d_n}_{1,n} \to \sum_{d_n=1}^{D^2} E^{d_n}_{1,n} \otimes
 E^{d_n',d_n}_{2,n}.
 \ee
This last tensor has more (right and left) indices than the
original one. Thus, every time we apply the MPO $U_k$ to a MPS
$|U_{k-1}\rangle$ the number of indices increases, and thus the
problem soon becomes intractable. Now we introduce a numerical
algorithm inspired by DMRG to numerically determine ${\rm
F}_2(E_{O_{h,c}})$ and to overcome this problem.

Given an unnormalized MPS $|\psi_A\rangle$ parameterized by
$D\times D$ matrices $\{A_k^s\}$, the goal is to find another MPS
$|\psi_B\rangle$, parameterized by $D_f\times D_f$ matrices
$\{B^s_i\}$, where $D_f< D$ is a prescribed number. This has to be
done such that $K:=\||\psi_A\rangle-|\psi_B\rangle\|^2$ is
minimal, i.e., such that that $|\psi_B\rangle$ gives the best
approximation to $|\psi_A\rangle$. We have developed an algorithm
that achieves this task in an iterative way. The key insight is
that $K$ is quadratic in all components of the matrices
$\{B^s_k\}$, and hence if all these matrices are fixed except one
of them (say $B^s_j$) $K$ is quadratic in the components of
$B^s_j$; the optimal choice for $B^s_j$ thus amounts to solving a
trivial system of linear equations. Having done that, one moves to
the next site $j+1$, fixes all other ones and repeats the same
procedure. After a few sweeps back and forth the optimal MPS is
found. Note that the error in the approximation is exactly known
and if it becomes too large one can always increase $D_f$; in all
relevant situations we encountered the error could be made very
small even with moderate $D_f$. The same reasoning holds for MPS
defined with periodic instead of open boundary conditions. In this
latter case considered here, one can further simplify the system
of linear equations by performing a singular value decomposition
of $B^s_j$ and keeping only one of the unitary matrices at each
step, analogously as one does in DMRG.

Thus, in order to evaluate an arbitrary expectation value we first
determine the MPS $|\tilde U_2\rangle$ which is the closest to
$U_2|U_1\rangle$ but with a fixed dimensions $D_f$ of the
corresponding matrices. Then, we determine $|\tilde U_3\rangle$,
which is the closest to $U_3|\tilde U_2\rangle$, and continue in
this vein until we finally determine $\langle
\Psi|O|\Psi\rangle\simeq \langle U_N|\tilde U_{N-1}\rangle$.
Interestingly enough, this method to calculate expectation values
and to determine optimal approximations to MPS can be adapted to
develop very efficient algorithms to determine the ground states
of 2D Hamiltonians and the time evolution of PEPS by extending
DMRG and the time evolution schemes to 2D.

Let us start with an algorithm to determine the ground state of a
Hamiltonian with short range interactions on a square $N_h\times
N_v$ lattice. The goal is to determine the PEPS (\ref{repres2D})
with a given dimension $D$ which minimizes the energy. Following
\cite{VPC04}, the idea is to iteratively optimize the tensors
$A_{h,c}^s$ one by one while fixing all the other ones. The
crucial observation is the fact that the exact energy of
$|\psi\rangle$ (and also its normalization) is a quadratic
function of the components of the tensor $A_{\bar h,\bar c}^s$ to
be optimized, which we write as a vector $x$; hence the energy can
be expressed in terms of an effective Hamiltonian:
 \be
E=\frac{x^\dagger \hat{H}_{eff} x}{x^\dagger \hat{N} x}
 \ee
The denominator takes the normalization of the state into account.
This expression can readily be minimized as it is equivalent to a
generalized eigenvalue problem. It turns out that $\hat{H}_{eff}$
and $\hat{N}$ can be efficiently evaluated by the methods
described above. In the case of $\hat{N}$, the MPS $|U_1\rangle$
(\ref{MPS2}) constructed from $E_{\openone_{h,1}}$ can be
propagated up to the row $\bar{v}-1$ with the technique outlined
before. Similarly, the last row $\langle U_{N_v}|$ (\ref{MPS3})
can be propagated up to row $\bar{v}+1$. The tensors
$E_{\openone_{h\bar{v}}}$ can now be contracted with these two MPS
from $h=1..\bar{h}-1$, and similarly from $h=N_h..\bar{h}+1$. The
remaining tensor has 4 (double) indices from which one can readily
determine $\hat{N}$. $\hat{H}_{eff}$ can be determined in an
analogous way, but here the procedure has to be repeated for every
term in the Hamiltonian (i.e. in the order of $2N_hN_v$ times in
the case of nearest neighbor interactions). Thus both $\hat{N}$
and $\hat{H}_{eff}$ can be calculated efficiently. Therefore the
optimal $A^s_{\bar{h}\bar{v}}$ can be determined, and one can
proceed with the following projector, iterating the procedure
until convergence.

In a practical implementation one can save much time by storing
appropriate tensors, implementing the algorithm in a parallel way,
doing sparse tensor multiplications, and making use of quantum
numbers and reflection symmetries. A more efficient variant can
also be constructed by an iterative procedure which resembles the
infinite-size DMRG-algorithm, where new rows are inserted in the
middle of the lattice.

Let us next move to describe how time-evolution can be simulated
on a PEPS. We will assume that the Hamiltonian only couples
nearest neighbors, although more general settings can be
considered. The simplest scheme would work by optimally mapping a
given PEPS to another PEPS after an infinitesimal time-step
$\openone-iH\delta t$. It can readily be checked that, up to first
order of $\delta t$, the action of this operator is to map a TPS
with auxiliary dimension $D$ onto a new TPS with dimension of the
auxiliary bonds $nD$; here $n$ represents the minimal number of
terms needed to express the couplings as tensor products of local
operators plus 1 (e.g. $n=2$ for the Ising interaction and $n=4$
for the Heisenberg interaction). In analogy to the method
introduced above for MPS, one can approximate this new PEPS with
another one having again bonds of auxiliary dimension $D$. The
algorithm to achieve this is a direct generalization of the method
introduced to reduce the $D$ of MPS: again several sweeps over all
projectors have to be done, and the only difference is that at
each step correlation functions of a PEPS have to be calculated
instead of correlations function of a MPS. This can be done using
the methods introduced before. Of course there are again many
possibilities to boost the accuracy and to reduce the
computational cost of such an implementation, such as using the
Trotter decomposition as in \cite{VidalDaleyWhite} and then using
the sweeps to optimize the state. This algorithm can also be used
to solve finite temperature or dissipation problems by extending
the ideas of \cite{Zwolak} and \cite{VGC04}.

Let us now illustrate our methods with an example. We consider a
2D lattice of spin $1/2$-particles where nearest neighbors
interact via the antiferromagnetic Heisenberg interaction with
coupling constant $J=1$. We use the time--evolution algorithm for
evolving the PEPS in imaginary time; in this way we illustrate
both the fact that the new formalism allows us to find ground
states as well as to describe time-evolution. We implemented the
algorithm as follows: we start with a completely separable state
$|\psi_0\rangle$ in which the spins are rotated by an angle
$\pi/16$ with respect to the previous one, and which can trivially
be written as a PEPS. Using the Trotter decomposition, we divide
each time step into 4 parts in which each spin is only interacting
with one neighbor; as we are considering the Heisenberg
interaction, the dimension $D$ between the 2 interacting spins
gets multiplied by a factor of 4. Let us parameterize this new
PEPS with the corresponding tensors $B_{h,v}^s$. After each of
these substeps, we want to reduce the dimension again to the
original one giving rise to the PEPS $C_{h,v}^s$ that optimally
approximates the exact $B^s_{h,v}$. This is done in an iterative
way, row by row, until convergence. Fixing all rows but one, the
problem of finding the optimal projectors in this row is
equivalent to the problem of approximating a MPS with another one
with lower dimension (the physical dimension of the MPS is the
product of the bonds going up and down), which can on itself done
in an iterative way as outlined above. Note that the computational
cost of the algorithm is polynomial in $N$ and $D$.

\begin{figure}[t]
  \centering
    \resizebox{\linewidth}{!}{\includegraphics{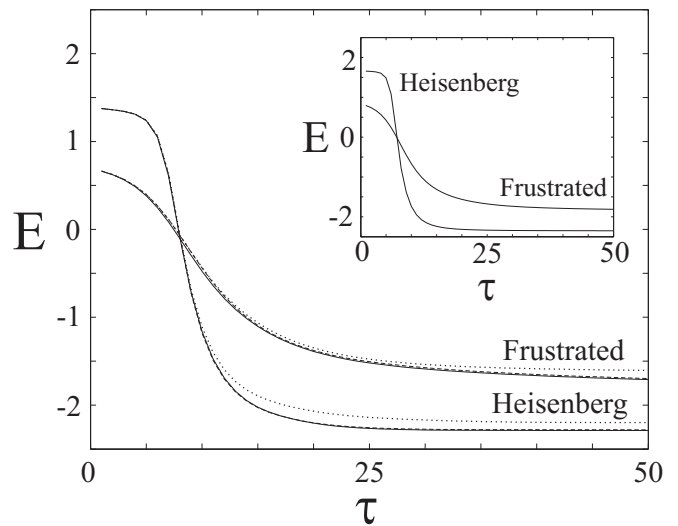}}
  \caption{Imaginary time evolution with the Heisenberg and a
  frustrated Heisenberg interaction on a $4\times 4$ lattice,
  and $D=2,D_f=16$ (dotted) and $D=3,D_f=35$ (dashed);
  the $D=3$ results are almost indistinguishable from the exact ones (full line).
  The insert presents the evolution for $D=2$ on a
  $10\times 10$ lattice.}
  \label{Fig2w}
\end{figure}

We have first considered a $4\times 4$ lattice on which the
imaginary time evolution can be determined exactly. In Fig.\ 2, we
plotted the exact evolution versus the one where the evolution is
approximated variationally within the PEPS with bonds of dimension
$D=2,3$ ($D=4$ cannot be distinguished from the exact result). We
used the same Trotter approximation for the exact and variational
simulations with $\delta t=-3i/100$. It is remarkable that even
for $D=3$ we obtain a very good approximations, both regarding
time evolution and ground state energy. The algorithm clearly
converges to the ground state, and the difference between the
exact ground state energy and the one obtained with our scalable
algorithm rapidly decreases with $D$\cite{fnD}; more specifically,
$1-E_{var}/E_{exact}$ is given by $.35,.02;.004;0.0008$ for
$D=1,2;3;4$ (note that the trivial situation $D=1$ corresponds to
the N\'eel state). We also repeated the same simulation but with a
frustrated Heisenberg Hamiltonian, obtained by making 1 out of
every 4 interactions on each spin ferromagnetic instead of
antiferromagnetic. Again very good agreement with the exact
results is obtained; note that the energy converges more slowly
due to the fact that the energy gap is smaller. The insert of
Fig.\ (\ref{Fig2w}) presents some simulation results for the
imaginary time evolution for a square $10\times 10$ lattice for
both the Heisenberg antiferromagnet and the frustrated case. The
convergence is again very fast, and increasing $D$ from 2 to 3
(not shown in the plot) allows us to find a better value for the
energy of the ground state. Note that we can easily handle larger
systems and, using the appropriate numerical techniques,
eventually increase the value of $D$.

In conclusion, we have introduced the class of PEPS and showed how
they arise naturally in the context of constructing variational
ground states for spin Hamiltonians on higher dimensional
lattices. We presented an efficient algorithm for calculating
correlation functions, which leads to scalable variational methods
for finding ground states and for describing their real or
imaginary time evolution. Interestingly, the methods described
also apply in the case of different geometries, of evolution in
the presence of dissipation, and for finding finite-T states. It
is also possible to identify quite generic classes of PEPS for
which 2-point correlation functions can be calculated analytically
\cite{inprep}. We also note that the concept of PEPS could be very
useful for the description of 2-dimensional transport problems, as
the PEPS generalize the matrix product states which proved to be
very useful in the 1-D case \cite{Derrida}.

We thank M. A. Martin-Delgado for his insights on DMRG and MPS.
Work supported by the DFG (SFB 631), european projects (IST and
RTN), and the Kompetenznetzwerk der Bayerischen Staatsregierung
Quanteninformation.


\begin{thebibliography}{99}

\bibitem{AKLT} I. Affleck {\it et al.}, Commun. Math. Phys. {\bf 115},
  477 (1988).

\bibitem{White} S.R. White, Phys. Rev. Lett. {\bf 69}, 2863 (1992).

\bibitem{VidalDaleyWhite} G. Vidal, Phys. Rev. Lett. 91, 147902 (2003);
A. Daley \emph{et al.}, J.Stat.Mech.: Theor.Exp. P04005 (2004); S.R. White and
A.E. Feiguin, cond-mat/0403310.

\bibitem{Zwolak} M. Zwolak and G. Vidal, cond-mat/0406440.

\bibitem{VGC04}  F. Verstraete, J.J. Garcia-Ripoll and J. I. Cirac,
cond-mat/0406426.

\bibitem{Fannes} M. Fannes, B. Nachtergaele and R.F. Werner, Comm.
Math. Phys. {\bf 144}, 443 (1992).

\bibitem{Ostlund} S. Ostlund and S. Rommer, Phys. Rev. Lett. {\bf 75}, 3537 (1995).

\bibitem{Dukelsky} J. Dukelsky {\it et al.}, Europhys. Lett. {\bf
    43}, 457 (1997).

\bibitem{VPC04} F. Verstraete, D. Porras and J.I. Cirac, cond-mat/0404706.

\bibitem{MC} D.M. Ceperley and B.J. Alder, Phys. Rev. Lett. {\bf 45}, 566 (1980).

\bibitem{Nishino04}  Y. Hieida, K. Okunishi and Y. Akutsu, New J. Phys. {\bf 1}, 7 (1999);
K. Okunishi and T. Nishino, Prog. Teor. Phys. {\bf 103}, 541 (2000); T. Nishino
\emph{et al.}, Nucl. Phys. B {\bf 575}, 504 (2000); Y. Nishio \emph{et al.},
cond-mat/0401115.

\bibitem{Niggeman} H. Niggeman, A. Kl\"umper and J. Zittartz, Z. Phys. B {\bf
104}, 103 (1997).

\bibitem{Martindelgado} M.A. Mart\'in-Delgado, M. Roncaglia and G. Sierra, Phys.
Rev. B {\bf 64}, 075117 (2001).

\bibitem{VC03} F. Verstraete and J.I. Cirac, quant-ph/0311130.

\bibitem{Vidallog} G. Vidal \emph{et al.},  Phys.Rev.Lett. {\bf 90}, 227902 (2003).

\bibitem{Xiang} S. Liang and H. Pang,  Phys. Rev. B {\bf 49}, 9214 (1994);
S.R. White, Phys. Rev. Lett. {\bf 77}, 3633 (1996); T. Xiang, J. Lou, and Z. Su,
Phys. Rev. B {\bf 64}, 104414 (2001).

\bibitem{fnD} The number of variational parameters scales as $D^4$, whereas
for 1D MPS it does as $D^2$. Thus, modest values of $D$ already
allow us to approximate the local 2-body density operators (and
hence the energy) to a good accuracy.

\bibitem{inprep} In Preparation

\bibitem{Derrida} B. Derrida \emph{et al.}, J. Phys. A: Math. Gen. {\bf 26}, 1493 (1993).

\end{thebibliography}
\end{document}